\documentclass{elsart}
\usepackage{natbib}
\usepackage{graphics}
\usepackage{longtable}

\usepackage{psfig,epsfig}
\usepackage{booktabs}

\newcounter{bla}                            
\newenvironment{refnummer}{%
\list{[\arabic{bla}]}%
{\usecounter{bla}%
 \setlength{\itemindent}{0pt}%
 \setlength{\topsep}{0pt}%
 \setlength{\itemsep}{0pt}%
 \setlength{\labelsep}{2pt}%
 \setlength{\listparindent}{0pt}%
 \settowidth{\labelwidth}{[9]}%
 \setlength{\leftmargin}{\labelwidth}%
 \addtolength{\leftmargin}{\labelsep}%
 \setlength{\rightmargin}{0pt}}}
 {\endlist}

\begin{document}
\newcommand{\xbeq}{\begin{eqnarray}} \newcommand{\xeeq}{\end{eqnarray}}

\runauthor{}
\begin{frontmatter}
  \title{Debris Disk\\ Radiative Transfer Simulation Tool (DDS)}
  \author[MPIA,Caltech]{S.\ Wolf},
  \author[Caltech]{L.\ A.\ Hillenbrand},

  \address[MPIA]{Max Planck Institute for Astronomy, K\"onigstuhl 17, 69117
	Heidelberg, Germany, swolf@mpia.de}

  \address[Caltech]{California Institute of Technology, 1201 E California Blvd, 
	Mail Code 105-24, Pasadena CA 91125, USA}

\begin{abstract}
A WWW interface for the simulation of spectral energy distributions of optically thin
dust configurations with an embedded radiative source is presented.
The density distribution, radiative source, and dust parameters can be
selected either 
from an internal database or defined by the user.
This tool is optimized for studying circumstellar debris disks where
large grains ($a_{\rm grain} \gg 1\mu$m) are expected to determine
the far-infrared through millimeter dust reemission spectral energy distribution.
The tool is available at {\tt http://aida28.mpia-hd.mpg.de/$\sim$swolf/dds}.
\end{abstract}


\vspace*{10mm}

\begin{flushleft}
CPC: 1.3  Radiative Transfer\\ [2ex]
  97.21.+a - Pre-main sequence objects, young stellar objects and protostars\\
  97.82.+k - Extrasolar planetary systems\\
\end{flushleft}

\vspace*{5mm}

\end{frontmatter}

\newpage
{\bf PROGRAM SUMMARY}
\bigskip

\begin{small}
\noindent
{\em Title of program:} Debris Disk Radiative Transfer Simulator (DDS)                  \\[10pt]
{\em Catalogue number:} (supplied by Elsevier)                \\[10pt]
{\em Program obtainable from:}\\ 
CPC Program Library, Queen's University
  of Belfast, N. Ireland (see application form in this issue) \\[10pt]
{\em Licensing provisions:} none  \\[10pt]  
{\em Computers:}\\
PC with Intel(R) XEON(TM) 2.80\,GHz processor
\\[10pt]
{\em Operating systems under which the program has been tested:}\\
SUSE Linux 9.1\\[10pt]
{\em Programming language used:}\\
Fortran 90 (for the main program; furthermore Perl, CGI and HTML) \\[10pt]
{\em Memory required to execute with typical data:}  $10^8$ words    \\[10pt]
{\em No. of bits in a word:} 8                            \\[10pt]
{\em No. of lines in distributed program, including test data, 
etc.:}\\
9264\\[10pt]
{\em Keywords:} \\[10pt]
{\em Nature of physical problem}\\  
Simulation of scattered light and thermal reemission in arbitrary optically
dust distributions with spherical, homogeneous grains where the dust parameters
(optical properties, sublimation temperature, grain size) and SED of 
the illuminating/heating radiative source can be arbitrarily defined
(example application: Wolf \& Hillenbrand~2003).
The program is optimized for studying circumstellar debris disks where
large grains (i.e., with large size parameters) are expected to determine
the far-infrared through millimeter dust reemission spectral energy distribution.
   \\[10pt]
{\em Method of solution}\\  
Calculation of the dust temperature distribution and dust reemission and scattering
spectrum in the optically thin limit.
   \\[10pt]
{\em Restrictions on the complexity of the problem}\\  
1) The approach to calculate dust temperatures and dust reemission spectra
is only valid in the optically thin regime. The validity of this constraint
is verified for each model during the runtime of the code.\\
2) The relative abundances of different grains can be arbitrarily chosen,
but must be constant outside the dust sublimation region., i.e.,
the shape of the (arbitrary) radial dust density distribution outside
the dust sublimation region is the same for all grain sizes and chemistries.\\
3) The size of upload files (such as the dust density distribution,
optical data of the dust grains, stellar spectral energy distribution, etc.)
is limited (see http://aida28.mpia-hd.mpg.de/$\sim$swolf/dds/ for current file size limits).
However, the resulting limitation to the complexity of possible model definitions
is marginal only.
   \\[10pt]
{\em Typical running time}\\   
3\,sec - 30\,min (depending on the complexity of the model)
   \\[10pt] 
{\em Unusual features of the program}\\ 
The program as provided through the CPC Program Library is equipped with
an HTML user interface. It is installed and available at\\
{\tt http://aida28.mpia-hd.mpg.de/$\sim$swolf/dds}.
   \\[10pt]
{\em References}
\begin{refnummer} 
\item Wolf, S., Hillenbrand, L.A., 2003, Astroph. Journal, 596, 603
\end{refnummer}

\end{small}

\newpage
{\bf LONG WRITE-UP}

\section{Introduction}

The {\em Debris Disk Radiative Transfer Simulation Tool} (DDS) 
was developed with the aim to provide a flexible tool
for the simulation of spectral energy distributions (SEDs) of circumstellar debris disks.
Debris disks are solar system-sized dust disks with micron-sized grains produced 
as by-products of collisions between asteroid-like bodies being left over from 
the planet formation process. In the case of our solar system, the debris
of Jupiter-family short-period comets and colliding asteroids represents the dominant 
sources of zodiacal dust located between Mars and Jupiter.  A second belt of dust
is located beyond the orbit of Neptune (see, e.g., Dermott et al.~1992, Liou, Dermott, 
\& Xu~1995).  Besides the solar system, optical to mid-infrared images of $\beta$~Pic
(see, e.g., Kalas \& Jewitt~1995; Weinberger, Becklin, \& Zuckerman~2003)
and submillimeter images of Vega, Fomalhaut, and $\epsilon$~Eri 
(Holland et al.~1998; Greaves et al.~1998) have revealed spatially resolved debris 
disks which were first inferred from observations of infrared flux excesses above 
photospheric values with IRAS. 
The mass of small grains in debris disks and therefore the thermal dust reemission 
from these disks is expected to be much smaller than in the case of T\,Tauri disks.
For this reason, only a very limited sample of observations exists so far. 
However, because of the high sensitivity of the mid-infrared detectors aboard the 
{\em Spitzer Space Telescope}
followed by the {\em Stratospheric Observatory for Infrared Astronomy} ({\em SOFIA}), 
a substantial increase in the total number and in the specific information about debris 
disks is expected (c.f. Meyer~2002). 

The range of applications of the DDS, however, extends
far beyond debris disks (or similar, e.g. disk-like structures) allowing to derive
the reemission and scattered light SED for {\em any} three-dimensional 
optically thin dust distribution:
In the optically thin limit only the radial distance $r$ between
each individual grain of a dust distribution determines this grain's contribution
to the net SED. Thus, any arbitrarily shaped configuration described, e.g. by
a three-dimensional denisty distribution
$\rho(x,y,z)$ can be reduced to an equivalent one-dimensional distribution $\rho(r)$
in respect of its resulting thermal reemission and scattered light SED
($r$ is the radial distance from the central star / heating source).
The DDS is therefore best characterized as a 
{\em tool
for the simulation of scattered light and thermal reemission in arbitrary optically
thin dust distributions with spherical, homogeneous grains where the dust parameters
(optical properties, sublimation temperature, grain size) and SED of 
the illuminating/heating radiative source can be arbitrarily defined}.

\newpage
\section{The method}\label{method}

The solution of the radiative transfer problem implemented in the DDS is based
on three assumptions:
\begin{enumerate}
\item The dust configuration is optically thin along any line of sight for both
  the stellar as well as for the reemission radiation from the dust,
\item The dust grains are compact spheres with a homogeneous chemical structure, and
\item The dust grains are large enough to be in thermal equilibrium with
  the ambient radiation field.
\end{enumerate}
In the optically thin limit, each dust grain is heated by direct stellar radiation only. 
Thus, the dust grain temperature is 
a function of the optical parameters of the grains, the incident stellar radiation, and the 
distance $r$ from the star. In this case the radiative transfer equation has a simple solution 
which allows one to derive the distance from
the star at which the dust has a certain temperature. Let
\begin{equation}\label{eq_dist1} 
  L_{\lambda, \rm s}         = 
  4\pi R_{\rm s}^2                         \pi B_{\lambda}(T_{\rm s})
\end{equation}
be the monochromatic luminosity of the star (radius $R_{\rm s}$, effective temperature 
$T_{\rm s}$) at wavelength $\lambda$ and
\begin{equation}\label{eq_dist2}
  L_{\lambda, \rm g}^{\rm abs} = L_{\rm s}        Q_{\lambda}^{\rm abs} \frac{\pi a^2}{4\pi r^2}
  \hspace*{1cm}{\rm and}
\end{equation}
\begin{equation}\label{eq_dist3}
  L_{\lambda, \rm g}^{\rm emi} = 
  4\pi a^2 Q_{\lambda}^{\rm abs} \pi B_{\lambda}(T_{\rm g})
\end{equation}
be the absorbed and reemitted luminosity of a dust grain with radius $a$ and resulting 
temperature $T_{\rm g}$
at the (unknown) radial distance $r$ from the star. Using the constraint of energy conservation
\begin{equation}\label{eq_dist4}
  \int_{0}^{\infty} L_{\lambda, \rm g}^{\rm emi} d\lambda = 
  \int_{0}^{\infty} L_{\lambda, \rm g}^{\rm abs} d\lambda
\end{equation}
one derives the distance of the grain from the star as
\begin{equation}\label{eq_dist5}
  r(T_{\rm g}) = \frac{R_{\rm s}}{2} 
    \sqrt{
    \frac{\int_0^{\infty} Q_{\lambda}^{\rm abs}(a) B_{\lambda}(T_{\rm s}) d\lambda}
	 {\int_0^{\infty} Q_{\lambda}^{\rm abs}(a) B_{\lambda}(T_{\rm g}) d\lambda}
	 }.
\end{equation}
If the dust sublimation temperature is known, Eq.~\ref{eq_dist5} also allows one 
to estimate the sublimation radius $r_{\rm sub}$ for each dust component 
in the shell (characterized 
by the grain radius and chemical composition).
The flux of the light scattered by a single dust grain amounts to
\begin{equation}\label{eq_scatt}
  L_{\lambda, \rm g}^{\rm sca} = L_{\lambda, {\rm s}}A Q_{\lambda}^{\rm sca} 
  \left(
  \frac{a}{2r}
  \right)^2,
\end{equation}
where $A$ is the dust grain's albedo.
The net spectral energy distribution results from a simple summation of the reemitted 
and scattered light contributions from all grains.  

The interaction of the stellar radiation field with the dust grains -
characterized by the efficiency factors $Q_{\lambda}^{\rm abs}$ and $Q_{\lambda}^{\rm sca}$ 
(and thus the albedo $A$) - is described by Mie scattering theory in the DDS. 
The Mie scattering function is calculated using the numerical 
solution for the estimation of the Mie scattering coefficients published 
by Wolf \& Voshchinnikov (2004; see also Voshchinnikov~2004), 
which achieves accurate results both in the small as well as 
in the --~arbitrarily~-- large size parameter regime.

\newpage
\section{Numerical Implementation}

\begin{figure}[ht]
    \bigskip
    \begin{center}
    \resizebox{\hsize}{!}{\includegraphics{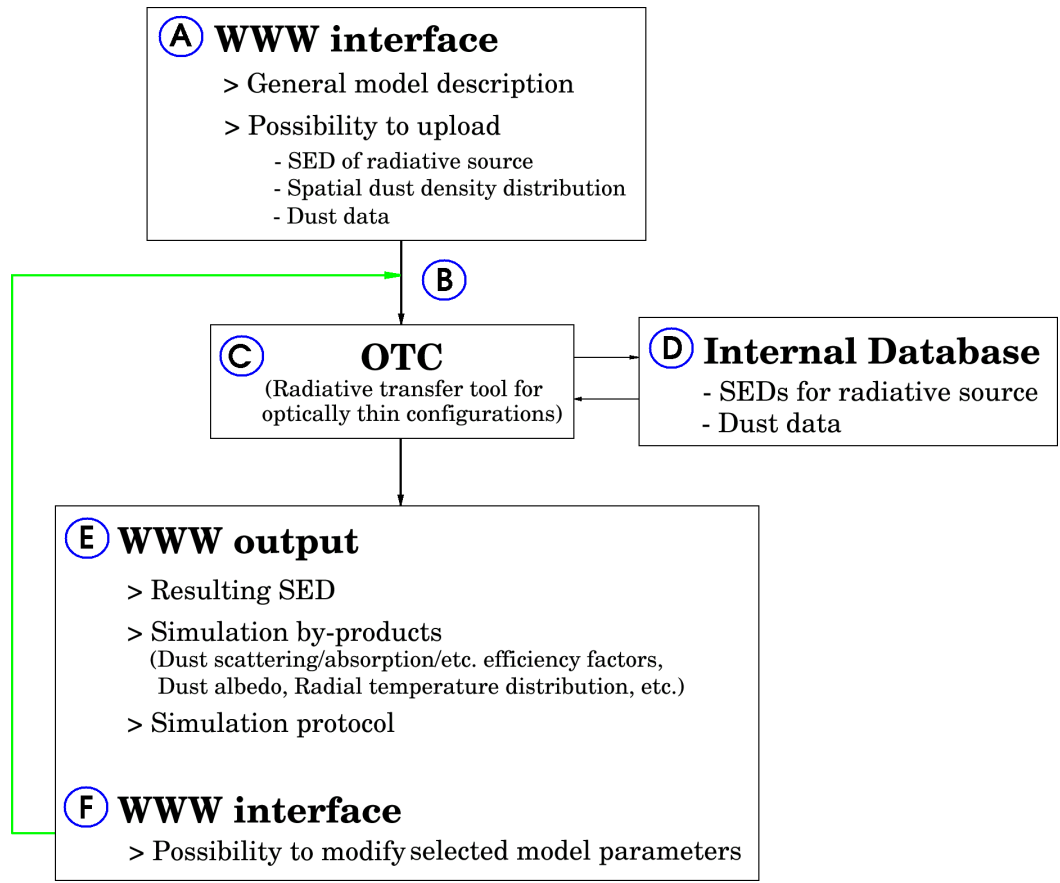}} 
    \end{center}
    \medskip
    \caption{DDS data flowchart.}
    \label{dds-flow}
    \bigskip
\end{figure}
In this section the general scheme and main features of the DDS are described.
For this reason a simplified chart symbolizing the data flow is given in Fig.~\ref{dds-flow}.
The single units, symbolized by letters {\tt A} - {\tt F} in Fig.~\ref{dds-flow}, 
solve following tasks:

\begin{figure}[ht]
    \begin{center}
    \resizebox{0.2\vsize}{!}{\includegraphics{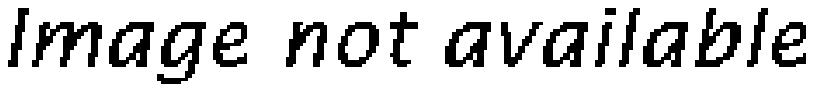}}
    \end{center}
    \medskip
    \caption{DDS: Input mask. {\bf Figure available in the complete article - see: http://aida28.mpia-hd.mpg.de/$\sim$swolf/dds/doc/swolf$\_$dds.pdf}}
    \label{dds-input}
\end{figure}

\noindent
{\bf A}:\\
WWW interface: Model setup (see Fig.~\ref{dds-input}).\\
Each debris disk model to be investigated with the DDS
(or dust configuration model in general) is defined by
\begin{enumerate}
\item 	The SED of the radiative source which is given
	\begin{enumerate}
	\item 	analytically as a blackbody radiator characterized by its temperature
		and total luminosity,
	\item	by an internally defined SED (e.g.\ the solar SED), or
	\item  	by the user; the format of the upload-files containing the SED
		is given in Sect.~\ref{upl-sed}.
	\end{enumerate}

\item 	The inner and outer radius of the disk.
	If an uploaded density distribution (see 3.\ below) is considered, the inner
	and outer radius defined in that file are used.
	If a fixed inner radius  is chosen which turns out to be within the sublimation
	region of a particular dust species (defined by its grain size and chemical 
	composition), the sublimation radius is used for this dust species instead.

\item 	The disk density distribution, which is either
	\begin{enumerate}
	\item	described analytically (by the radial density profile and opening
		angle of the disk), or
	\item 	provided by the user in form of an upload-file (see Sect.~\ref{upl-den}
		for the file format).
	\end{enumerate}

\item 	The disk mass, i.e.\ the total mass of all dust grains,

\item 	The dust grain size distribution given by the minimum and maximum grain size
	and an exponent that describes the slope of the distribution,

\item 	The relative abundances of the chemical components. Beside a menu of predefined
	chemical components the possibility to upload further components is provided
	(see Sect.~\ref{upl-dust} for the file structure).
	These abundances represent either relative mass densities per volume element
	or relative number densities of dust grains.

\item 	The specification of the observable SED to be simulated (wavelength distribution).
\end{enumerate}

\noindent
{\bf B}:\\
The handing over of the input parameters and file upload are managed by PERL and CGI scripts. 
The input parameters are stored in a file which is processed by unit C (see below).

\noindent
{\bf C}:\\
The OTC is a radiative transfer tool optimized for 
\underline{o}ptically 
\underline{t}hin 
\underline{c}onfigurations.
It is written in Fortran~90 and solves following tasks:
\begin{enumerate}
\item 	Evaluation of the input data:
	\begin{enumerate}
	\item 	Data type verification for each input value,
	\item 	Test of the model integrity, and
	\item 	Test of specific upper / lower limits of input parameter values.\\
	\end{enumerate}
	
\item 	Calculation of the SED:
	\begin{enumerate}
	\item	Definition of the wavelength range and wavelengths at which the dust
		absorption (and based on this the dust temperature distribution)
		shall be calculated.
		In the case of a blackbody radiative source, this wavelength range is derived
		under consideration of the blackbody temperature. A fixed number of
		wavelengths ($\sim 500$) is then distributed logarithmically equidistantly
		within this range. In the case of uploaded SEDs for the radiative source
		or the choice of one of the SEDs from the internal database (see unit~D below),
		the given wavelengths / wavelength range are used.

	\item	Calculation of the dust absorption efficiencies at the wavelengths
		of stellar emission (for dust absorption only) 
		and the user-defined observing wavelengths
		for each grain size~($a_i$) and chemical composition~($\xi_j$).

	\item	Calculation of the radial distances $r(T,a_i,\xi_j)$ of the dust grains
		to the heating source that correspond to predefined temperatures
		(see Eq.~\ref{eq_dist5}; 2.73K $\le T \le $ sublimation temperature).

	\item	If the density distribution is provided on a grid, the array 
		$r(T,a,\xi)$ is interpolated 
	        in order to provide the identical radial spatial resolution as the uploaded
		density distribution.

	\item	If required: Calculation of the scattering efficiency of the dust grains
		at the observing wavelengths.
	
	\item	Calculation of the relative net contribution of each individual dust species
		outside the corresponding sublimation radius.

	\item	Summation over the net contributions and normalization of the total flux
		based on the total dust mass in the model.\\
	\end{enumerate}

\item	Creation of all output files:
	\begin{enumerate}
	\item 	for the user: final results
	  (resulting SED, radial temperature distribution, etc.),
	\item	internally: 
		intermediate results, such as the relative contributions of each dust species
		to the net SED; these files are required in unit~F (see below).
	\end{enumerate}
\end{enumerate}

\noindent
{\bf D}:\\
A selection of astrophysically relevant dust 
data and SEDs of radiative 
sources\footnote{So far, only the solar SED is included 
(from measurements published by Labs \& Neckel~1968).
Further SEDs are planned to be included according to the response of the user community.}
is compiled in an internal database.
The optical data of following dust species have been included so far:
\begin{enumerate}
\item Silicates, oxides, and carbon configurations published 
  by Dorschner et al.~1995 and J\"ager et al.~1998 made available
  at\\
  {\tt http://www.astro.uni-jena.de/Laboratory/Database/odata.html} \\
  (Henning et al.~1999),
  and 
\item ``Astronomical silicate'' and graphite published by Weingartner \& \\ Draine~(2001). 
\end{enumerate}
Detailed references are given in unit~A.

\begin{figure}[ht]
    \begin{center}
    \resizebox{\hsize}{!}{\includegraphics{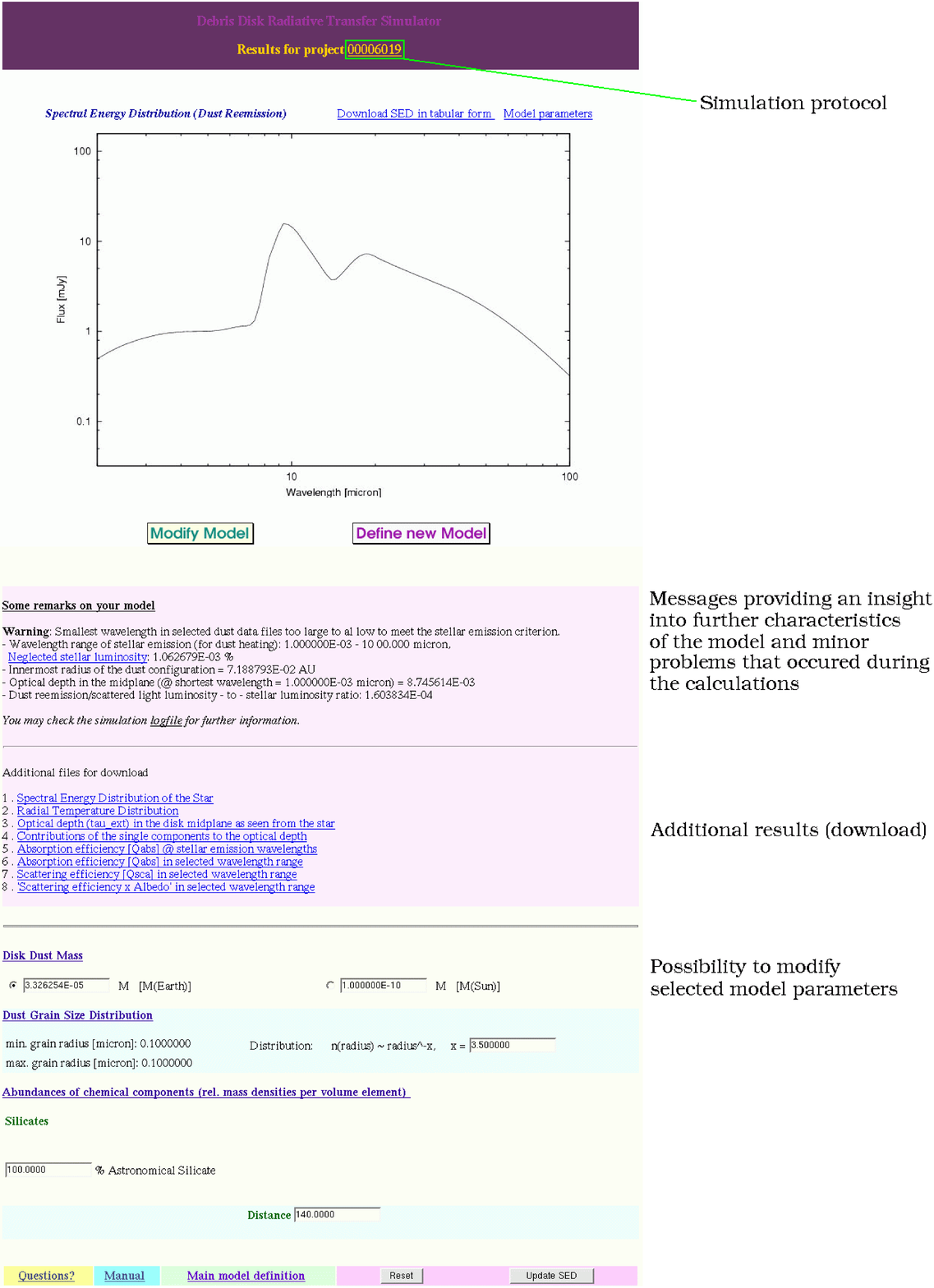}} 
    \medskip
    \caption{DDS: Example output.}
    \end{center}
    \label{dds-output}
\end{figure}
\noindent
{\bf E}:\\
Presentation of all results on a WWW page (see Fig.~\ref{dds-output} for an example).

\noindent
{\bf F}:\\
The WWW page with the results also allows to modify  selected model parameters, such as
\begin{enumerate}
\item	Disk mass,
\item 	Grain size distribution slope,
\item	Relative abundances of individual chemical components, and
\item 	Distance.
\end{enumerate}
Based on the intermediate results stored in unit C (step 3), a simple new weighting
of the relative flux contributions from the individual grains species
allows a quick calculation of the SED of the modified model.

\clearpage
\section{Performance of the code}

\begin{figure}[ht]
    \bigskip
    \begin{center}
      \resizebox{0.49\hsize}{!}{\includegraphics{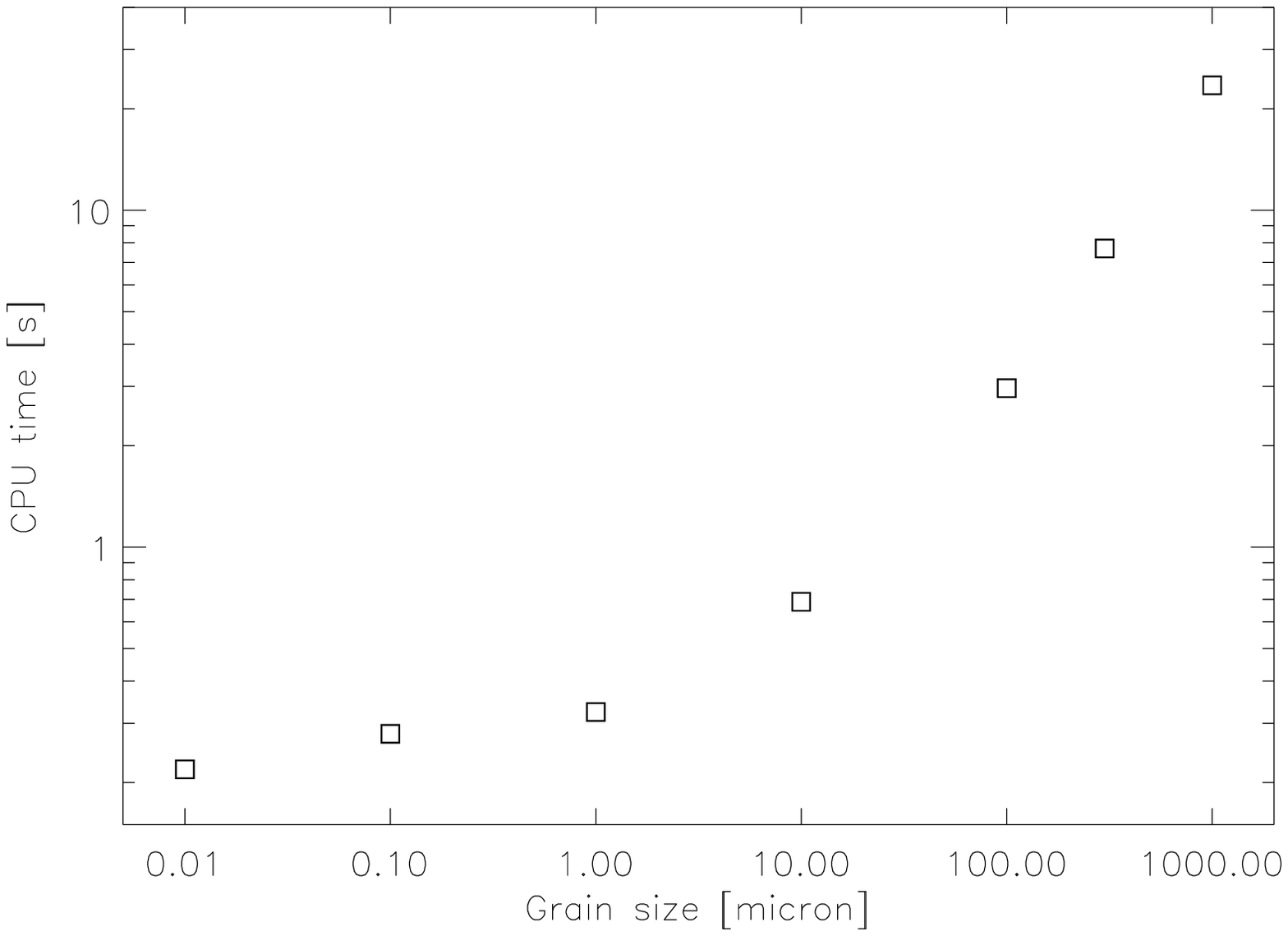}}
      \resizebox{0.49\hsize}{!}{\includegraphics{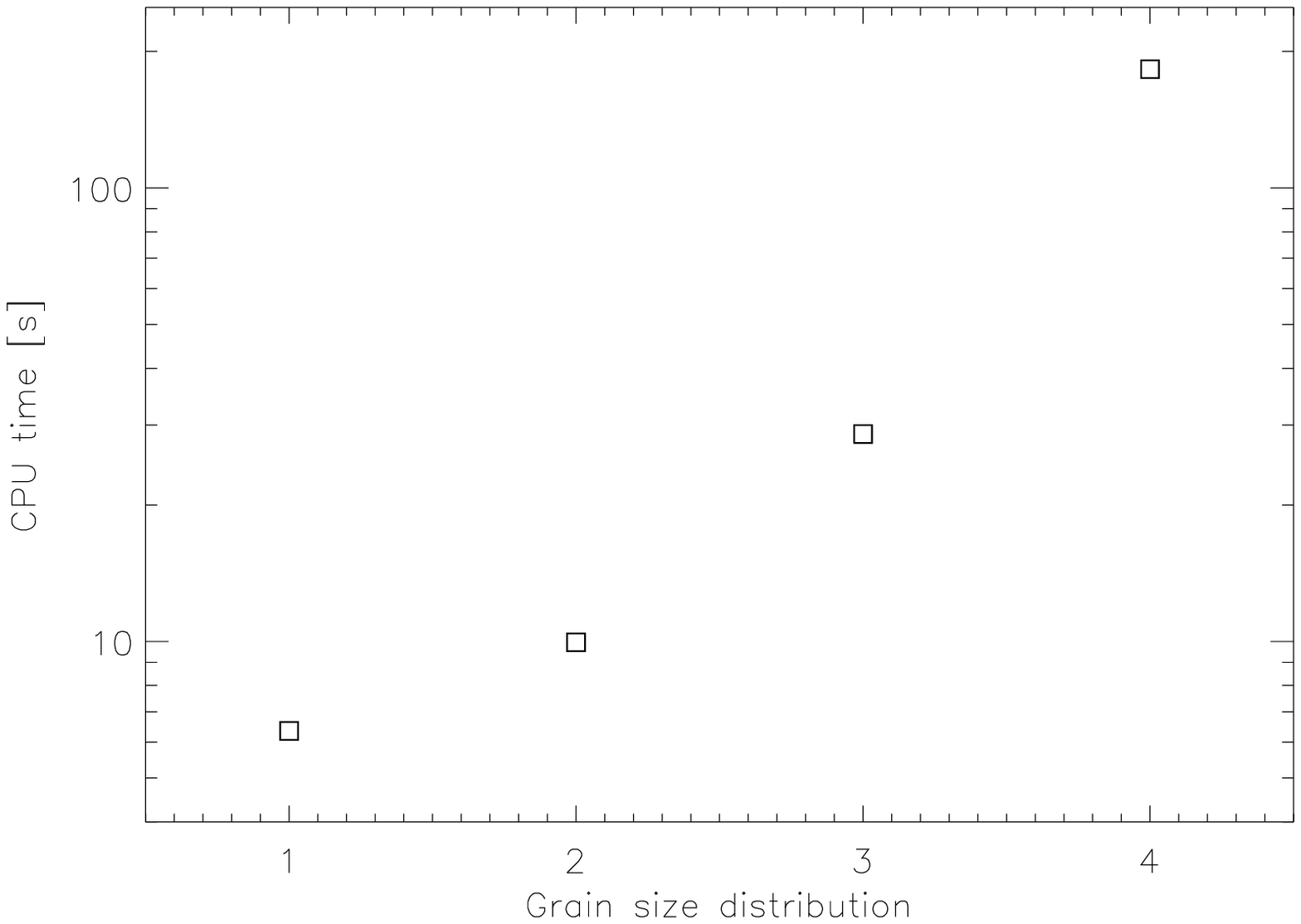}}
    \end{center}
    \medskip
    \caption{Performance of OTC/DDS: Run-time of the code
      as a function of grain size:
    {\bf [A]} Single grain sizes;
    {\bf [B]} Grain size distributions
    ({\tt 1}: 0.01-1$\mu$m,
    {\tt 2}: 0.1-10$\mu$m,
    {\tt 3}: 1-100$\mu$m,
    {\tt 4}: 10-1000$\mu$m).
    For this test the dust reemission has been calculated for 100
    wavelengths linearly distributed in the range 10-100$\mu$m.
    Otherwise, default settings have been used, such as 
    500 wavelengths to simulate the stellar emission.
    The CPU times have been estimated on an Intel(R) XEON(TM) 2.80\,GHz processor
    using the Intel Fortran~90 compiler for Linux.}
    \label{otc-perf}
    \bigskip
\end{figure}
The runtime of the code is mainly determined by the grain size (distribution) 
and number of wavelengths (a) at which absorption of the stellar light is modelled
(typically $\sim500$) and (b) at which the observable SED has to be calculated
(defined by the user).
The reason for this behaviour is that the dust grain parameters (efficiency factors,
albedo - see Sect.~\ref{method}) are calculated in each individual simulation
in order to provide maximum flexibility in the choice of the grain size distribution.
For a fixed model setup with a single grain size 
(defined by minimum grain size = maximum grain size)
the run-time of the code as a function of grain size is shown in Fig.~\ref{otc-perf}[A].
For selected grain size distributions 
(defined by: minimum grain size $<$ maximum grain size)
the run-times are shown in Fig.~\ref{otc-perf}[B]. In the latter case, 
32 individual grain sizes
logarithmically equidistantly distributed between the minimum and maximum grain size,
are considered for each chemical component.

\noindent
The OTC/DDS has been tested by comparing its results (radial temperature distribution,
reemission and scattered light SED) with those obtained with
the radiative transfer code MC3D (Wolf~2003). The results agree within the numerical
accuracy, i.e.\ both codes produce practically the same results.

\clearpage
\section{Acknowledgments}

S.W.\ was supported through the German Research Foundation 
(Emmy Noether Research Program WO 857/2-1), through the NASA grant NAG5-11645,
and through the SIRTF (Spitzer Space Telescope) legacy science program through 
an award issued by JPL/CIT under NASA contract 1407.
The program {\em gnuplot} [Copyright~(C) 1986-1993, 1998; Th.~Williams \& C.~Kelley]
is implemented in the DDS.
I wish to thank A. Moro-Mart\'{\i}n and J. Rodmann for their help to test the DDS
and all members of the FEPS team for valuable discussions.

\begin{appendix}
\section{Input file formats}

The SED of the illuminating and heating source, the dust density distribution, and
additional dust species can be uploaded in form of ASCII files.
These files must have tabular structures which are described 
in Sect.~\ref{upl-sed}-\ref{upl-obs}.
The file structures are described in an online help webpage, which
is connected via links to each of the individual upload sections.
Since the particular file structure of some of the input files may be changed over time,
only the links to the help pages are given in those cases.

\subsection{Stellar SED}\label{upl-sed}

The file structure for stellar SEDs is identical to internally predefined stellar SEDs
(see there for examples). It is documented at\\
{\em http://aida28.mpia-hd.mpg.de/$\sim$swolf/dds/dds-manual.html\#stellar\_sed\_upload}.

\subsection{Density Distribution}\label{upl-den}

\begin{table}[ht]
  \begin{tabular}{|p{65mm}|p{65mm}|}
    \hline
    \multicolumn{2}{|l|}{\# Header with remarks etc. (optional)}\\
    \multicolumn{2}{|l|}{\# $\ldots$}\\  \hline
    \multicolumn{2}{|l|}{Number radial grid points {\sl [integer]}} \\ \hline
    Radial distance $r$ from the source [AU] {\sl [float]}    
    &  Corresponding relative dust grain number density $n(r)$ {\sl [float]}\\
    $\ldots$                              &  $\ldots$\\\hline
  \end{tabular}
  \medskip
  \caption{File structure required for dust density distributions.}
  \label{tab-upl-dust}
  \bigskip
\end{table}

If $n_0(r,\theta,\phi)$ is the arbitrary (optically thin) density distribution, 
the radial density distribution $n(r)$ required for the simulation of the SED can be derived 
as follows (simple averaging):
\begin{equation}
  n(r) = 
  \frac{1}{4\pi}\int_0^{2\pi} \int_0^{\pi} n_0(r,\theta,\phi) \sin\theta \, d\theta\, d\phi.
\end{equation}
The structure of the file for density upload is documented in Tab.~\ref{tab-upl-dust}.

When providing the set of pairs $(r,n(r))$, the following guidelines should be considered:
\begin{enumerate}
  \item The distance between two subsequent radial points should decrease
    towards the heating source
    in order to allow the code (OTC) to resolve the increasing radial temperature gradient.
  \item The step size between two subsequent radial points has to be chosen smaller than 
    the typical size of structures in the density distribution in a particular distance $r$
    from the star in order to prevent skipping of local density enhancements, etc.\ 
    (for the simulation of the spectral energy distribution a linear increase/decrease
    of the density between two subsequent radial points is assumed).
\end{enumerate}

\noindent Further Remarks:
\begin{itemize}
  \item The quantity $n(r)$ represents the {\em relative} number of grains.
    The absolute number of grains is estimated internally based on
    the total mass of the dust configuration, specific dust density and grain size distribution.

  \item A radial grid with a logarithmic equidistant distribution of grid points 
    is a good choice in the case of "smooth" density distribution
    decreasing towards the radiative source. 
    In the case of a clumpy structure, however, the grid has to be adapted to the size 
    of the clumps in the (radial) regions where they are present. 
    The same applies if for instance circumstellar disks with gaps
    (due to planet-disk interaction) are considered.

  \item 
    If the dust density distribution results from an n-particle simulation,
    one should subdivide the model space in spherical shells centered on the star
    and estimate the mean relative number density of dust grains in each shell
    (total number of grains $N_i$ / volume of the shell $V_i$).
    Let $n(r_1)$ and $n(r_{100})$ be the number densities at the innermost and outermost shell 
    of a model subdivided into 100 shells. Let $r_1$ and $r_{100}$ be the corresponding
    mean radii of these shells. Then, the file prepared for upload of 
    the density distribution might have the following structure 
    (see also Fig.~\ref{den-exp-fig}):
    \begin{table}[ht]
      \begin{tabular}{|c|c|}
	\hline
	\multicolumn{2}{|l|}{\# Header with remarks etc. (optional)}\\\hline
	\multicolumn{2}{|l|}{102 -- number of subsequent lines}   \\ \hline
	Inner radius 	&  $n(r_1)$   \\\hline
	$r_1$ 	        &  $n(r_1)$   \\\hline
	$r_2$ 	        &  $n(r_2)$   \\\hline
	$\ldots$        &  $\ldots$ \\\hline
	$r_{99}$	&  $n(r_{99})$  \\\hline
	$r_{100}$ 	&  $n(r_{100})$ \\\hline
	Outer radius    &  $n(r_{100})$ \\\hline
      \end{tabular}
      \medskip
      \caption{Example upload-file for density distributions resulting from 
	n-particle simulations (see also Fig.~\ref{den-exp-fig}).}
      \label{tab-upl-den2}
      \bigskip
    \end{table}
    
\newpage
    \begin{figure}[ht]
      \begin{center}
	\resizebox{0.2\vsize}{!}{\includegraphics{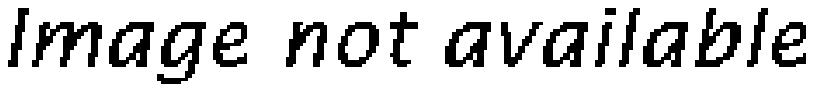}}
	\medskip
	\caption{Illustration of the subdivision of the model space 
	  in the case of density distributions resulting from n-particle simulations.
	The better the radial density distribution is sampled - especially in regions with
	high density gradients - the higher is the accuracy of the corresponding
	SED calculated with the DDS (see Sect.~\ref{upl-den}).
	{\bf Figure available in the complete article - see: http://aida28.mpia-hd.mpg.de/$\sim$swolf/dds/doc/swolf$\_$dds.pdf}}
      \end{center}
      \label{den-exp-fig}
    \end{figure}   
\end{itemize}
\newpage

\subsection{Dust Data}\label{upl-dust}

\begin{table}[ht]
  \begin{tabular}{|p{39mm}|p{45mm}|p{45mm}|}
    \hline
    \multicolumn{3}{|l|}{\# Header with remarks etc.\ (optional)}\\
    \multicolumn{3}{|l|}{\# $\ldots$}\\  \hline
    \multicolumn{3}{|l|}{Identifier (e.g. chemical composition) 
      {\sl [string]}} \\ \hline
    \multicolumn{3}{|l|}{Specific dust grain density in units of [g/cm$^3$] 
      {\sl [integer]}} \\ \hline
    \multicolumn{3}{|l|}{Sublimation temperature of the dust grains in units of [K]
      {\sl [float]}}   \\ \hline
    \multicolumn{3}{|l|}{Number of Wavelengths in the file 
      {\sl [integer]}} \\ \hline
    Wavelength [$\mu$m] {\sl [float]}     
    &  $Re$(refractive index) {\sl [float]} 
    &  $Im$(refractive index) {\sl [float]}\\
    $\ldots$                              
    &  $\ldots$          
    &  $\ldots$         \\\hline
  \end{tabular}
  \medskip
  \caption{File structure required for dust data upload-files.}
  \label{tab-upl-opt}
  \bigskip
\end{table}

A dust species of a particular chemical composition is defined by its specific
(material) density, its sublimation radius, and its wavelength-dependent complex refractive
index. A large database of laboratory measurements of astrophysically relevant
refractive indices is available at\\
{\tt http://www.astro.uni-jena.de/Laboratory/Database/odata.html}\\
(Henning et al.~1999).

The file structure for dust data is identical to those accessible through the DDS
(see there for examples). It is documented in Tab.~\ref{tab-upl-opt}.

\subsection{Observed SED}\label{upl-obs}

The DDS allows to upload observed SEDs in order to overlay them to the results
of the simulation.
The file structure for observed SEDs is identical to those created by the DDS
in order to allow an upload and overlay of simulated SEDs as well 
(see Sect.~\ref{down-sed}).
It is documented at 
{\em http://aida28.mpia-hd.mpg.de/$\sim$swolf/dds/dds-manual.html\#obs\_sed}.

\section{Output formats of selected files}

\subsection{Resulting SED}\label{down-sed}

\begin{table}[ht]
  \begin{tabular}{|p{40mm}|p{40mm}|p{20mm}|p{20mm}|}
    \hline
    \multicolumn{4}{|l|}{\# Header (model description)}\\ \hline
    Wavelength~[$\mu$m] {\sl [float]} &  
    Flux [mJy] {\sl [float]}          &
    0.0 &
    0.0 \\
    $\ldots$  &  $\ldots$  &  $\ldots$  &  $\ldots$  \\\hline
  \end{tabular}
  \medskip
  \caption{Structure of the output file with the calculated SED.}
  \label{tab-upl-res}
  \bigskip
\end{table}

The file structure for observed SEDs is identical to those created by the DDS
(in order to allow an upload of simulated SEDs as well).
It is documented in Tab.~\ref{tab-upl-res}.

\subsection{Radial temperature distribution}\label{down-tem}

\begin{table}[ht]
  \begin{tabular}{|p{75mm}|p{55mm}|}
    \hline
    \multicolumn{2}{|l|}{\# Header (model description)}\\ \hline
    Radial distance to the source $r$ [AU] {\sl [float]} &  
    Temperature [K] {\sl [float]} \\
    $\ldots$  &  $\ldots$ \\\hline
  \end{tabular}
  \medskip
  \caption{Structure of the output file with the radial temperature distribution.}
  \label{tab-upl-tem}
  \bigskip
\end{table}

The structure of the file with the radial temperature distribution is given
in Tab.~\ref{tab-upl-tem}. For a dust ensemble consisting of grains with different
sizes and chemical composition, the radial temperature profile is stored for 
each single grain species. Following algorithm to write out the data is implemented:
\\
\\
{\tt
for {\sl all chemical compositions}\\
\hspace*{5mm}for {\sl all grain sizes}\\
\hspace*{10mm}  write {\sl brief header containing the current\\
\hspace*{15mm}       chemical composition and grain size\\
\hspace*{15mm}       (starting with a ``\#'' sign)}\\
\hspace*{10mm}  for {\sl $r_{\rm sub} \le r \le$ outer radius}\\
\hspace*{15mm}      write {\sl r, T(r)}\\
\hspace*{10mm}  end for\\
\hspace*{5mm}end for\\
end for\\
}


\end{appendix}
\end{document}